%% file: ms.tex
\documentclass[12pt,preprint]{aastex}

\usepackage{amsmath,amssymb,latexsym,graphics}

\shorttitle{Microlensing in Q\,0957+561}
\shortauthors{Hainline et al.}

\newcommand{\kms}{\textrm{km\,s}^{-1}}
\newcommand{\mpc}{\textrm{Mpc}}
\newcommand{\msun}{M_{\sun}}

\begin{document}

\title{A New Microlensing Event in the Doubly-Imaged Quasar Q\,0957+561}

\author{Laura J. Hainline\altaffilmark{1}, Christopher W. Morgan\altaffilmark{1}, Joseph N. Beach\altaffilmark{1},
C. S. Kochanek\altaffilmark{2}, Hugh C. Harris\altaffilmark{3}, Trudy Tilleman\altaffilmark{3}, 
Ross Fadely\altaffilmark{4}, Emilio E. Falco\altaffilmark{5}, and Truong X. Le\altaffilmark{1}}

\altaffiltext{1}{Department of Physics, United States Naval Academy,
572C Holloway Rd, Annapolis, MD 21402, USA; hainline@usna.edu,
cmorgan@usna.edu,m110480@usna.edu,m113678@usna.edu}
\altaffiltext{2}{Department of Astronomy, The Ohio State University,
140 West 18th Ave, Columbus, OH 43210, USA; ckochanek@astronomy.ohio-state.edu}
\altaffiltext{3}{United States Naval Observatory, Flagstaff Station,
10391 West Naval Observatory Road, Flagstaff, AZ 86001-8521, USA;
hch@nofs.navy.mil, trudy@nofs.navy.mil}
\altaffiltext{4}{Department of Astronomy, Haverford College, 370 Lancaster Avenue, Haverford, PA 19041, USA;
rfadely@haverford.edu }
\altaffiltext{5}{Harvard-Smithsonian Center for Astrophysics, 60 Garden St, Cambridge, MA 02138, USA;
falco@cfa.harvard.edu}

\begin{abstract}

We present evidence for ultraviolet/optical microlensing in the gravitationally lensed
quasar Q\,0957+561.  We combine new measurements from our optical monitoring 
campaign at the United States Naval Observatory, Flagstaff (USNO) with measurements 
from the literature and find that the time-delay-corrected $r$ band flux ratio $m_{A} - m_{B}$
has increased by $\sim 0.1$\,magnitudes over a period of five 
years beginning in the fall of 2005.  We apply our Monte Carlo microlensing 
analysis procedure to the composite light curves, obtaining a measurement of 
the optical accretion disk size, $\log \{(r_{s}/\textrm{cm})[\cos (i)/0.5]^{1/2}\} = 16.2 \pm 0.5$, 
that is consistent with the quasar accretion disk size -- black hole mass relation.

\end{abstract}

\keywords{gravitational lensing: strong --- gravitational lensing: micro ---
                    accretion disks --- quasars: individual (Q\,0957+561)}

\section{INTRODUCTION}

The history of microlensing in the $z_{s}=1.405$ quasar Q\,0957+561 
(hereafter Q0957), the first confirmed gravitational lens system, is paradoxically 
simple yet complex.  The possibility of microlensing in this double-image, wide-separation 
($\sim 6\arcsec$) quasar was proposed by \citet{chang79} soon after its 
discovery by \citet{walsh79}.  Controversy over the time delay, currently accepted to be 
$\Delta t_{AB} = 417\,$days \citep[e.g.,][]{vanderriest89,schild90,kundic97,shalyapin08}, 
spawned numerous time variability studies of Q0957 at optical and 
radio wavelengths which produced a wealth of monitoring data.
Microlensing of the quasar on long time scales was soon established by a
change in the time delay-corrected magnitude difference of the two images of 
$\Delta (m_{A} - m_{B}) \sim 0.25$ over the course of five years, 
observed by \citet{vanderriest89} and confirmed by \citet{pelt98}.
However, after that initial event, occurring between 1982--1986, the
difference light curve (the difference between the light curve for
image A and the time-delay shifted light curve for image B) became nearly constant.  
Debate over the existence of measurable variability in the flux ratio percolated through the
literature, with some authors reporting variability in the flux ratio on the order of 
several hundredths of a magnitude over timescales of weeks, days, and hours,
while other groups simultaneously concluded that no microlensing with amplitude
$ | \Delta (m_{A} - m_{B}) |  \gtrsim 0.05$ was detected 
\citep[see, e.g.,][for summaries and references]{gilmerino01,schmidt10}.
As the dispute surrounding the existence of short-timescale microlensing
evolved, the absence of long-time scale microlensing became increasingly evident from the
difference light curve's steadiness over roughly 20 years.  The only report of
a possible long-time scale event after 1986, that of \citet{ovaldsen03} of 5\%-level
variability over 300~days, was not corroborated independently by \citet{wambsganss00} with data
taken over the same time period.  Studies as 
recent as \citet{shalyapin08} find no evidence for microlensing
in their data.

With the development of numerical simulation methods to analyze microlensing
observations \citep[e.g.,][]{kochanek04,bate08,blackburne11}, quasar
microlensing has meanwhile become a powerful tool with which to quantitatively study
the properties of lens galaxies and the structure of quasars.  Although quasars and
individual stars in lens galaxies cannot 
be resolved by conventional telescopes, by modeling the
amplitudes of microlensing fluctuations we can measure the size of the continuum
emission region, the masses and velocities of the stellar
microlenses, and the stellar-to-dark mass fraction in the lens galaxy 
\citep[e.g.,][]{kochanek04,pooley07,morgan08,mediavilla09,dai10}.
Through analysis of microlensing variability at multiple wavelengths, the
surface brightness profile of the quasar may be measured \citep[e.g.,][]{poindexter08,eigenbrod08}.
We are now moving into an era in which theoretical models of 
accretion disk structure can be tested with significant samples of lensed
quasars, resulting in the observational confirmation of a relation
between central black hole mass and accretion disk size \citep{morgan10},
as well as the temperature profile of the \citet{shakura73} thin accretion disk model 
\citep[e.g.,][]{poindexter08,anguita08,mosquera11}.

Since the central black hole in Q0957 lies at the high end of the quasar
black hole mass function \citep[$1.0\times 10^{9}\,\msun$;][]{assef11},
Q0957 provides a potentially interesting and important data point for testing
models of accretion disk structure.  However, the available analyses
of Q0957's only widely-acknowledged microlensing event from 1982--1986
pre-dated the use of the newer, highly quantitative microlensing
analysis techniques, and the qualitative methods employed in the published
analyses of the event produce either limits or
constraints too broad to be informative for current tests of
quasar structure and studies of the surface mass density of lens galaxies.
For example, \citet{pelt98} found that the 1982--1986 microlensing event
was consistent with a quasar radius of ``roughly''
$3\times 10^{15}\,\textrm{cm}$ and could be explained by microlens
masses down to $10^{-5}\,\msun$ with only a small fraction ($<5\%$) in solar
mass objects.  \citet{refsdal00} also analyzed the 1982--1986 microlensing
event and subsequent 8-year quiescent period in Q0957 using three different
simple lensing mass distribution models, obtaining rough constraints on the 
range of microlens masses of $10^{-6} < M/\msun < 5$ and an upper
limit on the size of the quasar's optical continuum region of 
$R < 10^{16}$\,cm.  Attempts to analyze the negligibly low microlensing
variability in Q0957's difference light curve with early numerical
simulations \citep{schmidt98,wambsganss00} could not 
constrain the quasar size.  Such studies could only rule out microlens masses
for a given quasar size, and in the end obtained results contradictory to 
analyses of the 1982--1986 microlensing event.  Thus, the lack
of unambiguous microlensing events in Q0957 in the relatively recent era of
computationally-intensive microlensing analysis has been unfortunate, preventing
improved determinations of the quasar's size and microlens mass
properties.

However, as the saying goes, ``good things come to those who wait.''
In this paper, we combine three new seasons of optical photometric 
monitoring of Q0957 with previously-published light curves to demonstrate the return 
of long-time scale uncorrelated variability in the light curves of Q0957.  We take advantage
of this new microlensing event to place the best constraints yet on the size
of the optical accretion disk and the mean mass of the microlenses using the 
Bayesian Monte Carlo analysis technique of \citet{kochanek04}.  In 
\S\ref{sec:obs_data} and \S\ref{sec:lens_model} we present our new monitoring observations of 
Q0957 and discuss the methods we use to model the uncorrelated variability in the light curves.  We
present our results in \S\ref{sec:results}.  Throughout our
discussion we assume a flat cosmology with $\Omega_{\textrm{M}}=0.3$, 
$\Omega_{\Lambda}=0.7$, and $H_{0}=70\,\kms~\mpc^{-1}$ \citep{hinshaw09}.

\section{OBSERVATIONAL DATA}\label{sec:obs_data}

We regularly monitor the flux of Q0957 A and B in the Sloan Digital Sky Survey $r$-band using the 
1.55-m Kaj Strand Astrometric Reflector at the United States Naval Observatory (USNO), 
Flagstaff Station, as part of the United States Naval Academy (USNA)/USNO
Lensed Quasar Monitoring Program.  Our program, which began in 2008, obtains three five-minute exposures
of the quasar per night, one night per week on average (weather permitting), using either
the Tek2K CCD camera ($0\farcs33\,\textrm{pixel}^{-1}$) or the
$2048 \times 4096$ EEV CCD camera ($0\farcs18\,\textrm{pixel}^{-1}$).
We present here data taken on 57 nights between 2008 March and 2011 June.  The 
median stellar FWHM (seeing) of the images in our data set is $1\farcs3$.

We measure the quasar image fluxes using relative photometry.  A detailed
discussion of our image analysis methods was presented in \citet{kochanek06},
so we only briefly summarize our procedure here.   We treat each quasar image
as a point source and model the point spread function with three nested 
elliptical Gaussian components, keeping the relative astrometry fixed to that
derived from \emph{Hubble Space Telescope} (\emph{HST}) images by 
\citet{keeton00} (see Table~\ref{tab:q0957_astrom}). 
The $z_{l}=0.356$ lens galaxy \citep{falco97} is modeled as a Gaussian approximation to a 
deVaucouleurs profile for which we fix the effective radius 
to that derived from the \emph{HST} images of Q0957.  The lens
galaxy's flux is held constant in all epochs; we determine the optimal 
value for the constant flux by repeatedly fitting all the images as a function of lens galaxy flux 
and finding the value which results in the lowest $\chi^{2}$ sum
over the entire data set.  We compare the flux of each quasar image 
to four reference stars, located at $(-59\farcs2, -27\farcs8)$, $(-61\farcs7, -109\farcs1)$,
$(-68\farcs5, -89\farcs8)$, and $(+111\farcs8, -127\farcs1)$ relative to 
Q0957 image A.   Because these reference stars are significantly
redder than the quasar images ($g - i \sim 0.8 - 0.9$ for the stars
and $g - i \sim 0.2$ for the quasar), and the two different detectors used
in our USNO program have slightly different spectral responses
across the $r$ band,
a small color term between the different detectors is introduced into 
our relative photometry measurements.  We determined this
color term by fitting the light curves of the quasar images obtained with
the EEV camera to the light curves obtained from the Tek2K camera,
yielding a magnitude offset of
$m_{\textrm{Tek2K}} - m_{\textrm{EEV}} = 0.048 \pm 0.002\,\textrm{mag}$.  We
apply this correction to the data from the EEV camera, and  
we include the contribution of the uncertainty in the color term, added in
quadrature, in the photometric errors of measurements taken with the EEV camera.

The measurements of Q0957 A and B from our USNO monitoring program
are listed in Table~\ref{tab:lightcurve}.  We note that we 
obtained data on two additional epochs
which have not been included in Table~\ref{tab:lightcurve}, bringing
the total number of observation epochs to 59. However, the images 
obtained on the two excluded dates were unusable due to high sky brightness
levels on one night and corruption of the quasar images by cosmic rays 
on the other.

\begin{figure}
\begin{center}
\plotone{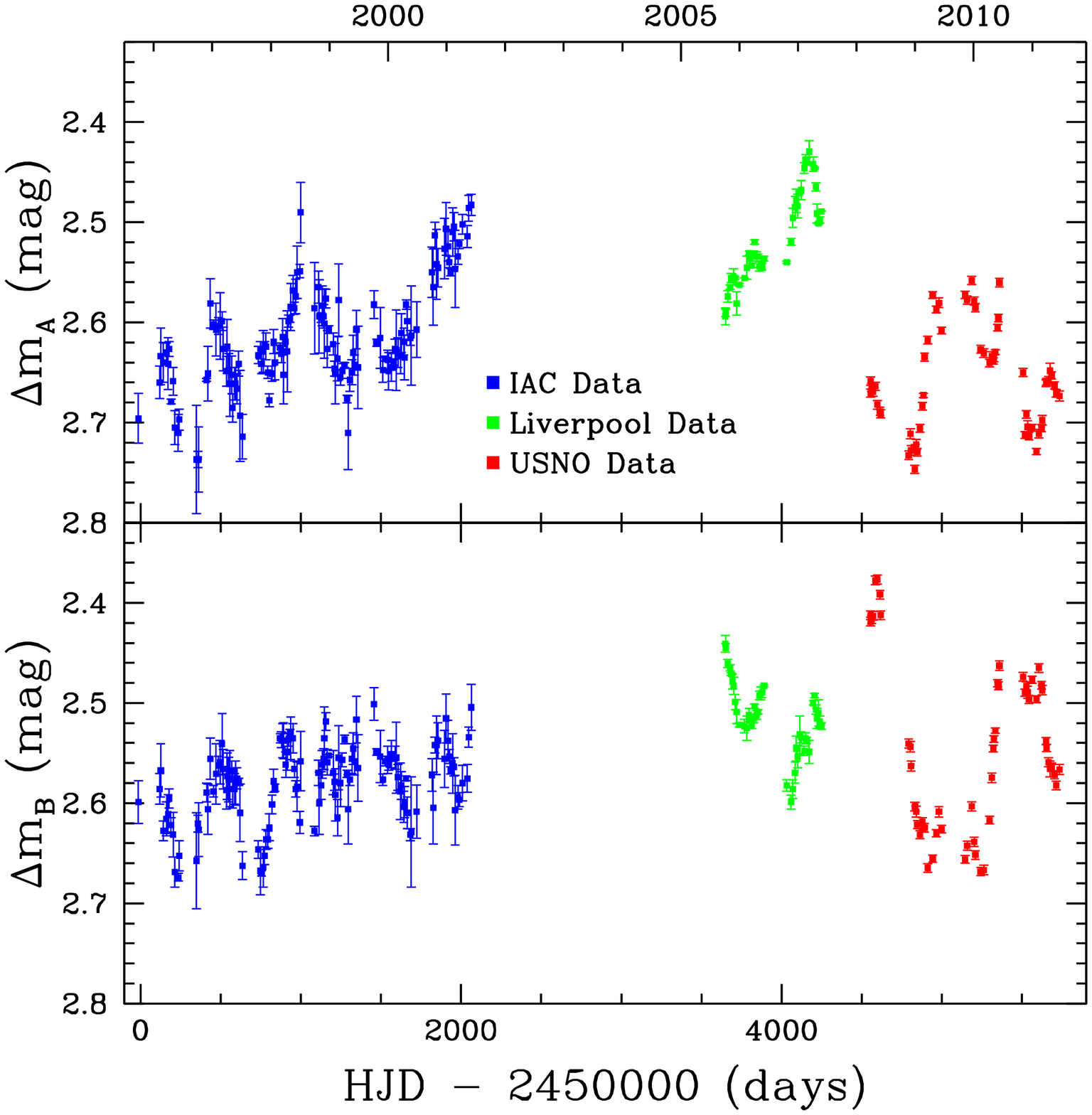}
\caption{Composite $r$-band light curve of Q0957 including historical measurements from
\citet[][IAC data]{serra99}, \citet[][IAC data]{oscoz01,oscoz02}, 
\citet[][Liverpool data]{shalyapin08}, and our new USNO data. 
The $\sim 1600$~day gap in the center of the light curve 
reflects an absence of published photometric monitoring for this system, but will
not prevent us from deriving useful constraints from our microlensing analysis.}\label{fig:comp_lc}
\end{center}
\end{figure}

In order to obtain a longer time baseline for our microlensing analysis, we
have supplemented our new USNO light curves with published monitoring data.  We 
use the $R$-band photometry from the 0.8-m telescope at the Instituto de Astrof\'{i}sica de Canarias'
(IAC) Teide Observatory spanning the years 1996--2001 from \citet{serra99}, \citet{oscoz01}, 
and \citet{oscoz02}, deriving the photometric zeropoint offset by 
comparison to concurrent $r$-band measurements of image A from
\citet[][$\Delta m_{\textrm{S99}-\textrm{K97}}=0.04\,\textrm{mag}$]{kundic97}; we neglect the small
color term difference between $r$ and $R$ because in our microlensing analysis
we use only the relative flux ratio and not the absolute flux.  
We also use the $r$-band monitoring data from \citet{shalyapin08}, 
spanning the years 2005--2007, and obtain the photometric offset ($14.455 \pm 0.018\,\textrm{mag}$) by 
analyzing their Liverpool Telescope/RATCam images\footnote{\citeauthor{shalyapin08}\
have made their monitoring images publicly available at the website
http://dc.zah.uni-heidelberg.de/liverpool/res/rawframes/q/form.} for five 
different nights with our image analysis procedure and averaging the 
offsets in magnitude between the published photometry and the values 
produced by our pipeline.  As the frequency of monitoring in those two data
sets was nightly, on average, and our monitoring program obtains observations with
a frequency closer to weekly, we have averaged the individual measurements into
seven-day bins within individual observing seasons, after removing obvious ($> 3\sigma$) 
outliers in the data set, to prevent those data from having a disproportionate
statistical weight relative to the new measurements from USNO.  The center of
each seven-day bin is defined as the mean HJD of the measurements included
in that bin.  To serve as 
errors on these seven-day averages, we compute the standard deviation of the individual 
data points included in each average.  We show the final composite light curves for 
Q0957A and B from all of the data sources in Figure~\ref{fig:comp_lc}.   Unfortunately,
we are unable to find any published monitoring data for Q0957 which falls in the gap 
in Figure~\ref{fig:comp_lc} between 2001--2005; while the absence of data in this 
date range will affect our microlensing analysis, we will still derive meaningful results.

We investigated including the two seasons of $r$-band monitoring from \citet{kundic97} from 
1995--1996 in our composite light curve for Q0957, which would have extended our time 
axis even further.  However, we chose to not include this data
in our analysis because the flux contribution to image B from the lens galaxy has
not been subtracted from the photometry in that data set.   Moreover, archival images from
this data set are no longer available at the Apache Point Observatory or any of
its operating consortium institutions, so it was not possible to carry out the
deconvolution of image B from the lens galaxy ourselves.

\section{ANALYSIS}\label{sec:lens_model}

\subsection{Difference Light Curves and Microlensing Signal}

To analyze the light curves of Q0957 for the presence of uncorrelated variability, 
we must first eliminate the variability intrinsic to the quasar source itself.   
We accomplish this by first shifting the light curve of image B by the system's 417-day 
time delay and then performing a linear interpolation of image B's shifted light 
curve to generate a set of photometric measurements at the same epochs of observation 
as those in image A's (unshifted) light curve. We discard any data points that 
were interpolated in the inter-season gaps.  Finally, we subtract from the light curve 
of image A the shifted light curve of image B, creating a time-delay-shifted 
difference light curve in which only the uncorrelated variability remains.\footnote{We
note that if we instead shift and interpolate the light curve of image A to the 
observational epochs of image B, no significant changes in the difference
light curve or analysis results are observed.}

\begin{figure}
\begin{center}
\plotone{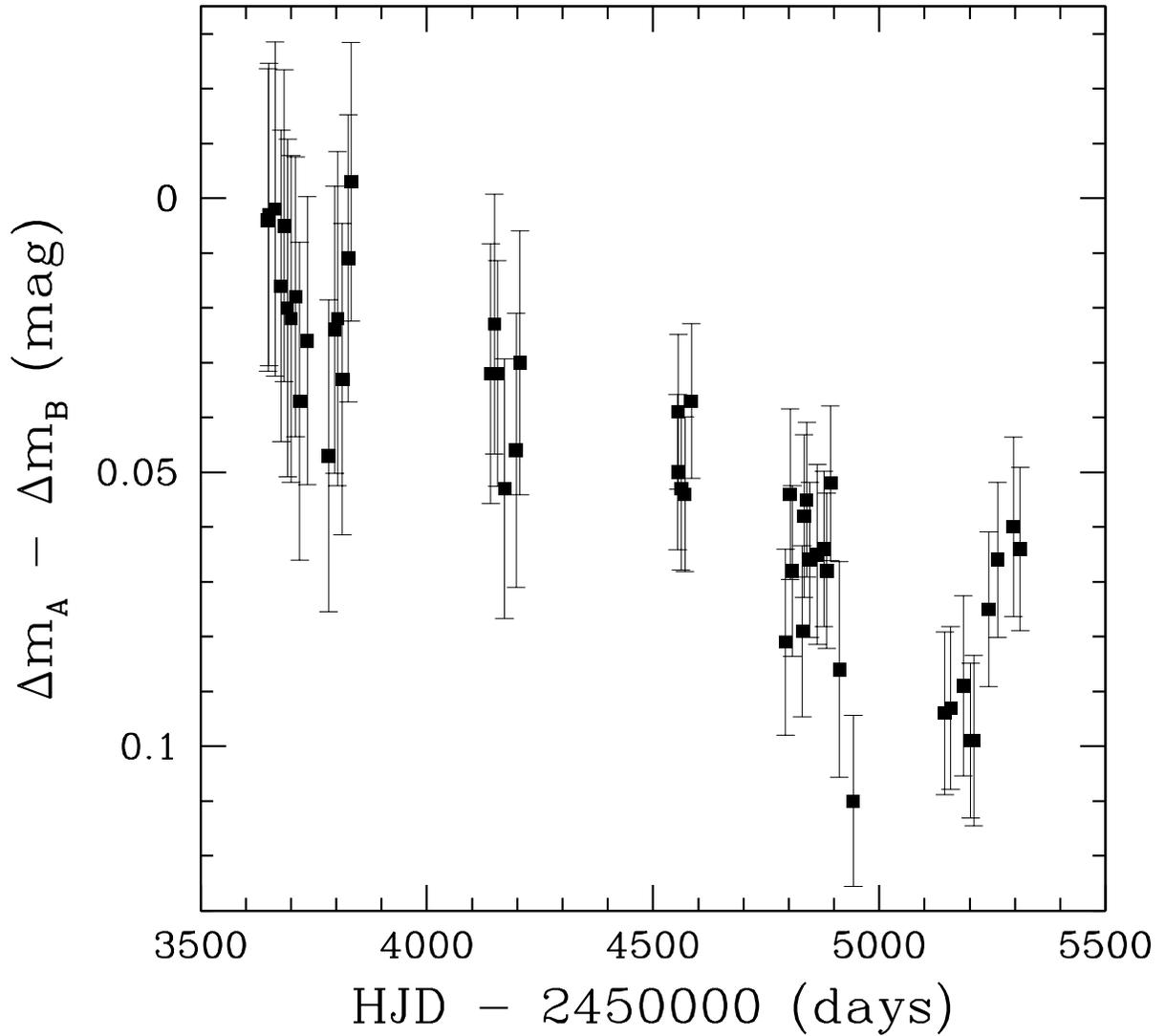}
\caption{Time-shifted difference light curve of Q0957 
from 2005--2010, displaying the slow increase in flux ratio of images A and B
over five years that is a clear signature of microlensing.  A simple linear
fit to the observed data points indicates an average change of
$\sim 0.02\,\textrm{mag\,yr}^{-1}$.}\label{fig:diff_lc}
\end{center}
\end{figure}

In Figure~\ref{fig:diff_lc} we present the time-delay-shifted difference light curve
for Q0957, focusing on the last six years of data.   We note in advance that 
the frequency of observations of Q0957 in the data sets we use does 
not permit us to confirm or refute the reported observations of intra-day microlensing 
variability \citep{colley03}.  Moreover, the size of the error bars on 
the measurements render insignificant hints of flux ratio variability on time 
scales of weeks or months, so we cannot examine reports of the short time scale
microlensing variability in Q0957 \citep{schild96,colley00,ovaldsen03}.
We do observe a slow but steady brightening of image B relative to image A  
in the $r$ band, beginning at the start of the \citet{shalyapin08} data near MHJD 3650
(in the frame of image A) and continuing to the limit of our USNO data set for image B
at MHJD 5311.  This clear signature of microlensing is evolving quite slowly, 
as indicated by the slope of a least-squares straight line fit to the difference 
light curve of $0.016\pm 0.006\,\textrm{mag\,yr}^{-1}$ (quoted uncertainty is $3\sigma$).   
Such a slow drift in flux ratio would not have been statistically
significant at the time \citeauthor{shalyapin08}\ published their data, if even noticeable, as they
could only constrain the flux ratio up to date MHJD 3831 in the frame of image A with their
data set; our monitoring data from USNO provide the extension in time necessary to confirm
that the change in flux ratio is significant in comparison to the observational uncertainties.
The duration of the complete microlensing event is presently unclear from the shape 
of the difference light curve, but the nearly five year span (after time delay correction)
so far implies the time scale will be at least that long.  We are thus fortunate to be
monitoring the first significant multi-year microlensing event in Q0957 in twenty years, 
and the first to be observed since the system's time delay was firmly established.  

\subsection{Quantitative Monte Carlo Microlensing Analysis}

The long time scale and low amplitude of the new microlensing variability are 
qualitatively consistent with the relatively large optical quasar size predicted by 
\citet{pelt98} and \citet{refsdal00} in their analyses of the 1982--1986 microlensing event 
and are not surprising given that the black hole is very massive.  
We can place new, improved quantitative constraints on the quasar size and the mass and 
velocity of the microlenses by analyzing the microlensing variability 
in our compiled light curve of Q0957 with the techniques described in \citet{kochanek04}.
The analysis method has three major components.  First, with the selection of
a model for the macroscopic (strong) lensing, a stellar mass function to 
describe the microlens mass distribution, and an accretion disk model,
we generate microlensing magnification patterns for
a range of mass contributions from the dark matter halo of the lens galaxy.
Next, we use a Monte Carlo method to generate large numbers of trial light 
curves from the magnification patterns for random source (quasar) trajectories,
and fit the simulated light curves to the full observed light curves (years 1995--2011).  
Finally, we perform a Bayesian statistical
analysis on the goodness-of-fit ($\chi^{2}$) statistics of the light curve fits to calculate
probability distributions for accretion disk size, effective source velocity, and mean 
microlens mass.  

To describe the macroscopic lensing in Q0957, we utilize the results from
the study of \citet{fadely10}, who develop detailed models of the quasar lens
using as constraints faint knots and structures visible in 
quasar- and lens-subtracted \emph{HST} images of the strong lensing region 
as well as the results of weak lensing analysis.  The lens potential in Q0957 has been 
notoriously difficult to model \citep[e.g.,][]{kochanek91,grogin96,bernstein99,keeton00}, at least in part
due to the two-image nature of the system and the presence of a galaxy
cluster surrounding the lens galaxy.  \citeauthor{fadely10}\ construct a series
of models containing an isophotal model of the lens galaxy as the stellar component, 
a concentric elliptical dark matter halo 
component representing the lens galaxy and the cluster halo, and a set
of general third-order terms in the Taylor series expansion of the potential from
the lens environment.  Here, we use
the convergence ($\kappa$) and shear ($\gamma$) of the
model from \citeauthor{fadely10}\ which lies at the peak of the posterior
probability distribution to generate the microlensing magnification patterns.  
So that we may marginalize over the uncertainty in the dark matter fraction 
in our analysis, we create a series of ten models in which the stellar component of the 
local convergence $\kappa_{\ast}$ in the vicinity of image B varies linearly 
in the range $0.1 < (\kappa_{\ast} /\kappa)_{B} < 1.0$.  Since image A 
is much farther from the lens galaxy, we constrain its stellar mass fraction 
$(\kappa_{\ast}/\kappa)_{A}$ such that the ratio 
$(\kappa_{\ast}/ \kappa)_{A} / (\kappa_{\ast}/\kappa)_{B}$ is held fixed 
to its value in the \citeauthor{fadely10} model $(\sim 0.2)$.
We note that the flux ratio favored by the \citeauthor{fadely10}\ models 
($f_{b}/f_{a} \approx 0.53$) is somewhat lower than 
what is actually observed in the optical \citep[1.1;][]{shalyapin08,goicoechea05}.

The magnification patterns are $4096 \times 4096$ images with an
outer scale of $20\langle R_{E} \rangle$, where $\langle R_{E}\rangle$ is the 
Einstein radius for the mean microlens mass $\langle M \rangle$ projected
into the source plane.  The outer dimension and pixel scale are chosen to
be sufficiently large to representatively sample the magnification patterns and
sufficiently small to resolve the accretion disk.  The stellar mass function we use to 
generate the population of microlenses for the patterns is a power law, 
$dN/dM \propto M^{-1.3}$, with a ratio of
maximum-to-minimum mass (dynamic range) of 50.  This function reasonably
approximates the Galactic disk mass function of \citet{gould00}; assuming a
different mass function \citep[such as that of][]{salpeterimf} will have a 
negligible effect on our results due to the other, larger sources
of uncertainty \citep[see, e.g.,][]{paczynski86,wyithe00}.  We create four independent magnification
patterns for each quasar image for all ten evenly spaced stellar mass fractions.

Because the effects of finite continuum source size are generally not negligible for
quasar microlensing, we convolve the magnification patterns with the surface 
brightness profile of the source for a grid of source sizes before computing 
trial light curves.   We model the accretion disk as a face-on, thin disk 
radiating as a blackbody with a power-law temperature profile $T \propto R^{3/4}$.
The scale radius of the disk, $r_{s}$, is defined
as the radius at which the disk temperature matches the rest-frame wavelength
of the filter used in our monitoring observations, $kT = hc/\lambda_{\textrm{rest}}$
(for $r$-band monitoring of Q0957, $\lambda_{\textrm{rest}} = 2593\,\textrm{\AA}$).
Our model matches the outer regions of the thin disk model of \citet{shakura73},
but we neglect the drop in temperature in the center due to the inner edge of the 
disk and the correction factor from general relativity to avoid introducing additional
parameters.  The effect of the simplification on our result for the accretion disk size
is small compared to the uncertainties in other parameters, as long as the disk is
significantly larger than the radius of the inner disk edge \citep{dai10}.  
We caution that because microlensing amplitude depends on the 
projected area of the source and not its shape, and because we assume
a face-on disk model, the disk radius we infer will be an effective radius defined
as the radius of a circle of the same projected area as the accretion disk.
The true radius will be $(\cos i)^{-1/2}$ times this effective radius, where $i$ is
the inclination of the disk.

We generate $10^{6}$ trial light curves for each of the 40 sets of 
magnification patterns.  In each realization
we randomly select an initial position and effective velocity for the source 
trajectory, assuming the values are independent and uniformly distributed.
As in \citet{kochanek04}, we neglect the motion of the stars within the lens
galaxy and describe the observer's motion as the projection of the CMB dipole velocity
onto the lens plane.  We then use Bayesian methods to analyze the
$\chi^{2}$ statistics of the many light curve fits and calculate likelihood functions 
for the quasar source size and velocity in Einstein units ($\hat{r}_{s}$ and $\hat{v}_{e}$)
since all results scale with the unknown mass of an average microlens
$\langle M \rangle$.   We estimate the true, unscaled physical 
source size $r_{s} = \hat{r}_{s}\langle M/\msun \rangle^{1/2}$ and mean 
microlens mass $\langle M \rangle$ by combining the probability distributions of 
the scaled source size, $P(\hat{r}_{s})$, and the scaled
effective velocity, $P(\hat{v}_{e})$, obtained from the Bayesian analysis 
with a prior probability function for the 
true, unscaled effective source velocity, $P(v_{e})$.  We construct $P(v_{e})$ using the
method described in \citet{kochanek04}, applying the measured stellar velocity dispersion for
the lens galaxy \citep[$\sigma_{\ast} = 288 \pm 9\,\kms$;][]{tonry99}, and 
obtaining the dispersion of the peculiar velocity distribution at the redshifts of
Q0957 and the lens galaxy from the power-law fits by \citet{mosquerakochanek11} 
to the peculiar velocity models of Tinker et al.\ (2011, in preparation).

\section{RESULTS AND DISCUSSION}\label{sec:results}

\begin{figure}
\begin{center}
\includegraphics*[angle=-90,scale=0.65,trim=0pt 12pt 0pt 0pt,clip]{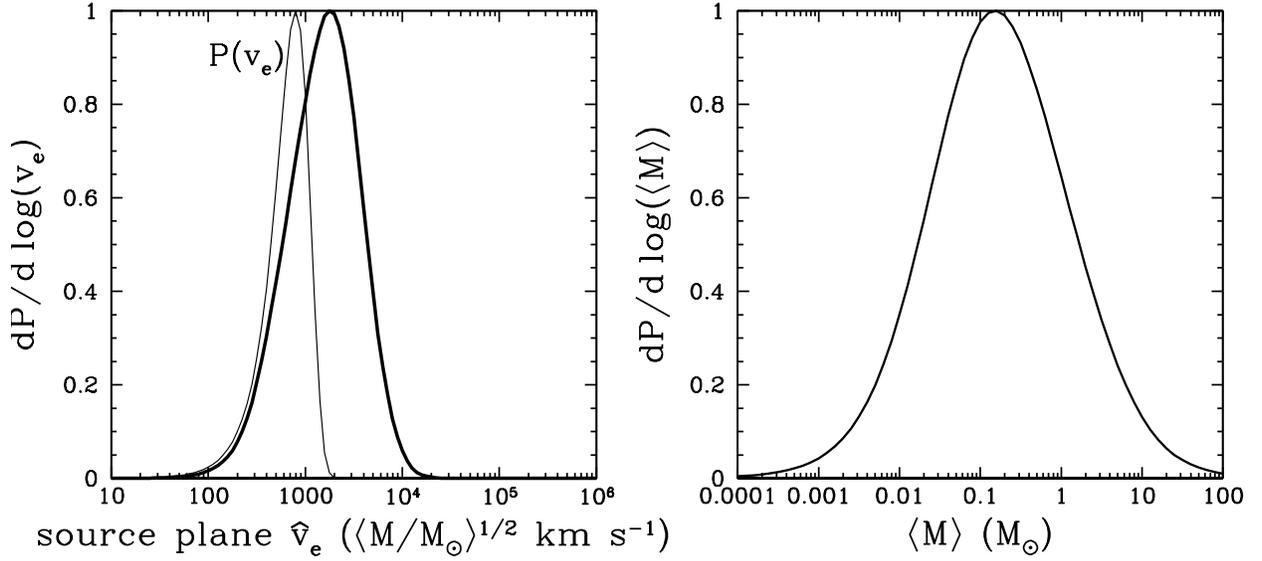}
\caption{\emph{Left panel}: Probability distribution for the effective
source velocity $\hat{v}_{e}$ for Q0957.  The heavy line is the scaled effective velocity distribution
in Einstein units, which has median $\hat{v}_{e} = 1600\,\kms$, while the light line indicates the prior probability
distribution for the true source velocity.  \emph{Right panel}: Probability distribution for the mean 
microlens mass $\langle M \rangle$ for the lens galaxy in the Q0957 system, which
has a median value of $\langle M \rangle = 0.2\,\msun$.  
$\langle M \rangle$ is calculated by combining the scaled effective source velocity
distribution with the prior probability distribution for the true source velocity.  The
width of the $\langle M \rangle$ distribution reflects the uncertainty in the 
effective velocity apparent in the left panel, since $\langle M \rangle$  depends on the 
inverse square of the scaled effective velocity $\hat{v}_{e}$.}\label{fig:vel_dist}
\end{center}
\end{figure}

\begin{figure}
\begin{center}
\plotone{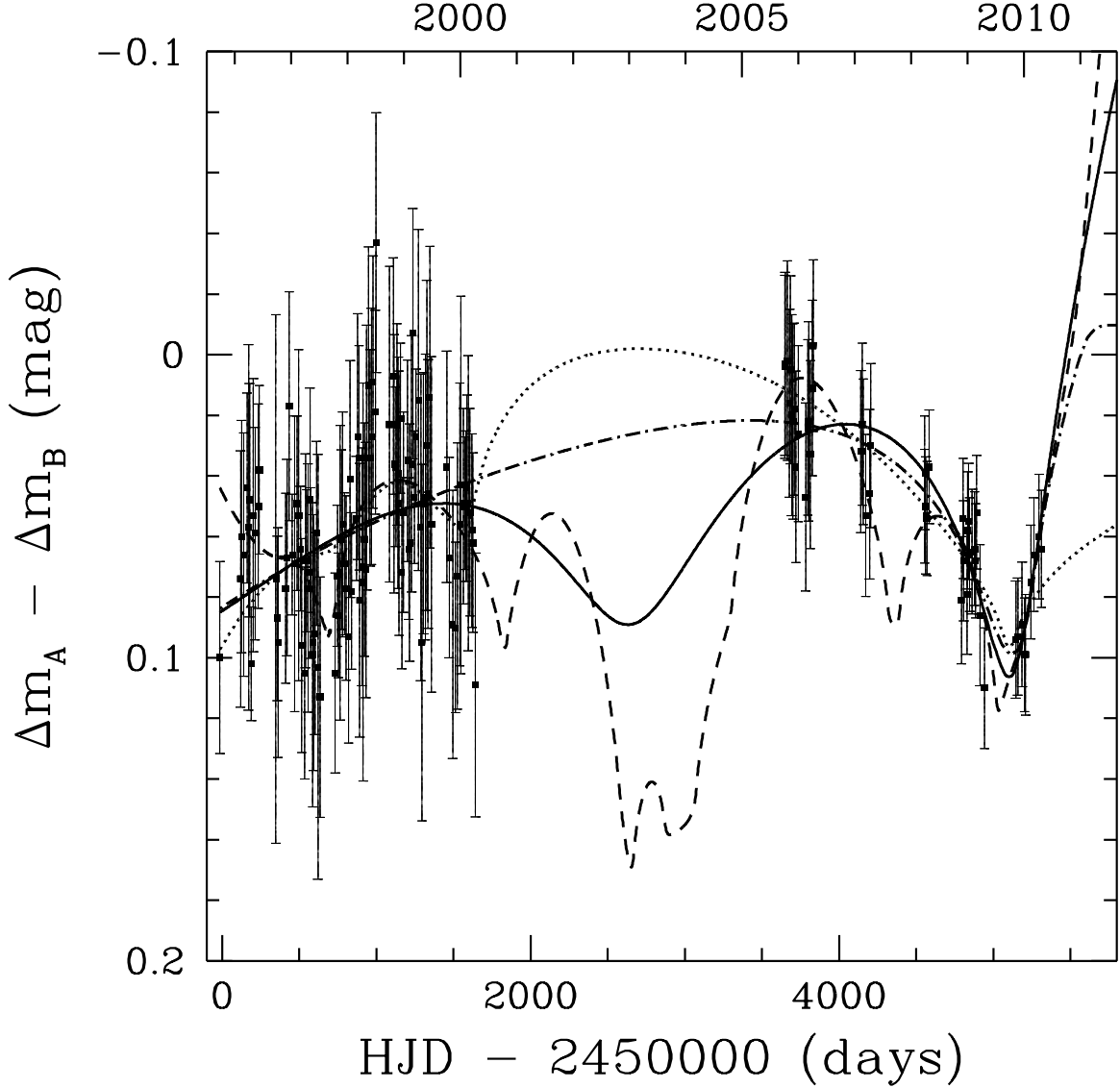}
\caption{Time-shifted difference light curve of Q0957 with a variety
of simulated light curves that are good fits to the observed data, illustrating
the ease with which the observed data are fit with simulated microlensing trajectories
for a variety of different physical parameters.
The different line types  correspond to simulations with $\kappa_{\ast}/\kappa = 0.1$
and $\hat{v}_{e} = 1468\,\kms$ (solid line), $\kappa_{\ast}/\kappa = 0.5$ and $\hat{v}_{e} = 561\,\kms$ 
(dotted line), $\kappa_{\ast}/\kappa = 0.8$ and $\hat{v}_{e} = 344\,\kms$ (dashed line),
and $\kappa_{\ast}/\kappa = 0.3$ and $\hat{v}_{e} = 2082\,\kms$ (dot-dashed line).}\label{fig:ml_model_fits}
\end{center}
\end{figure}

In Figure~\ref{fig:vel_dist} we show the probability distribution for the scaled effective
source plane (Einstein) velocity $\hat{v}_{e}$ resulting from the Bayesian analysis of our Monte Carlo
light curve simulations, along with the probability density for
the true source velocity, $P(v_{e})$, used to determine 
the mean microlens mass $\langle M \rangle$ and physical source size $r_{s}$.
The $\hat{v}_{e}$ distribution has a median of $1600\,\kms$ and is wide, 
with 68\% confidence range of $600\,\kms < \hat{v}_{e} < 3500\,\kms$.  The distribution is
also visibly asymmetric, with a substantial low-velocity tail.  The width and asymmetry indicate 
the wide range of feasible solutions to the microlensing variability in
Q0957 and reflect the ease with which 
the system's light curve was fit by the Monte Carlo code.  Because the Monte Carlo simulations
produce large numbers of acceptable fits and the $\chi^{2}$ per degree 
of freedom for the best fits tends to be quite low ($\sim 0.4$) even with insignificant (small) 
assumed values for the systematic errors of the flux ratio and photometry, we conclude that the
spread in the velocity distribution is a consequence of the low intrinsic variability of
Q0957 coupled with the very low amplitude of the microlensing signal, making the light curve
simple to reproduce.   An additional factor may be that the lack of observational constraints 
in the four-year gap in the compiled light curve permits considerable freedom 
in model behavior, as shown in Figure~\ref{fig:ml_model_fits}.   We tested this hypothesis
by conducting an experiment in which we forced the flux ratio in the 
gap to stay within $\pm 0.05$\,mag of its mean value in the dates bracketing the gap,
$(m_{A} - m_{B}) = 0.04 \pm 0.03\,\textrm{mag}$.  The experiment 
resulted in a minor narrowing of the $\hat{v}_{e}$ distribution, but the change 
was not significant enough to justify any firm conclusions about the influence of 
the gap on our results.

Because the mean microlens mass is proportional to
the inverse square of the scaled effective velocity ($\hat{v}_{e} = v_{e}\langle M/\msun \rangle^{-1/2}$),
the broad range in permitted velocities 
causes the mean mass to be poorly constrained, which is apparent in the probability 
distribution for $\langle M/\msun \rangle$ shown in the right panel of Figure~\ref{fig:vel_dist}.
The median of the $\langle M/\msun \rangle$ distribution is $0.2\,\msun$, with a
68\% confidence range of $0.02\,\msun < \langle M \rangle < 1.3\,\msun$.
Our estimate of the mean microlens mass is appropriate for stars in an old 
elliptical galaxy, and falls well within the most probable range
found by \citet{refsdal00} of $10^{-6} < M/\msun < 5$.   Our constraints
do not permit a dominant population of planetary and sub-planetary mass ($< 10^{-3}\,\msun$), compact 
microlenses in the lens galaxy, consistent with the results of \citet{schmidt98} and
\citet{wambsganss00}.  Our new constraints are also not consistent with the result of \citet{pelt98} for 
the 1982--1986 event, in which a significant population of stellar mass microlenses is ruled out. 
However, as our microlensing analysis technique is significantly more sophisticated 
than that used in earlier studies and does not rely on the assumption of
a single value for the source size or effective velocity to derive the microlens
mass, we consider our new result to be much more robust.

While we have attempted to constrain the stellar-to-dark
mass fraction $\kappa_{\ast}/\kappa$ in the lens galaxy of the Q0957 system, the
probability distribution resulting from our Bayesian analysis is uninformative, with no
strong peaks or trends favoring any particular $\kappa_{\ast}/\kappa$ value.
We suspect that the primary reason for this failure is that 
there is insufficient uncorrelated variability in the light curves of Q0957 
to constrain the dark matter fraction.  However, the addition of X-ray data to the 
microlensing analysis may provide stronger constraints, as it has in the case of
PG\,1115+080 \citep{morgan08,pooley09} and RXJ\,1131-1231 \citep{dai10}.

\begin{figure}
\begin{center}
\plotone{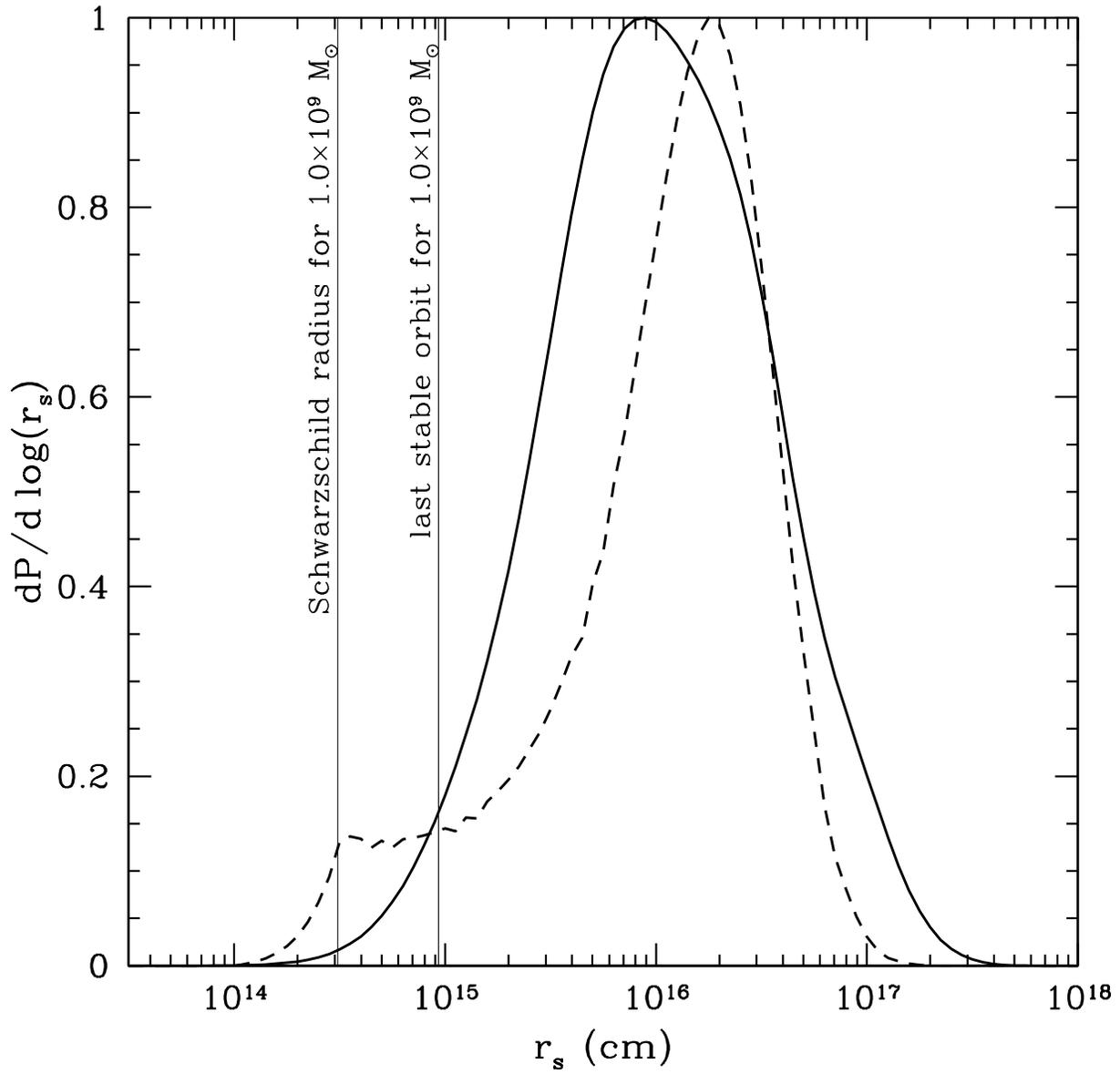}
\caption{Relative probability distribution for the physical thin disk scale size $r_{s}$ for Q0957.  
The solid line represents the probability distribution directly arising from the 
microlensing simulations, while the dashed line shows the result of imposing 
a prior on the mean microlens mass of $0.1 < \langle M/\msun \rangle < 1.0$.}\label{fig:size_dist}
\end{center}
\end{figure}

In Figure~\ref{fig:size_dist} we show the probability distribution for the
physical source size $r_{s}$ of the quasar's accretion disk in the observed-frame
$r$-band which results from our microlensing analysis.   The distribution as shown has
not been corrected for inclination $i$; however, in the text that follows all the numerical
quantities we discuss will be corrected by a factor of $(\cos i)^{-1/2}$ assuming an
inclination of $60\degr$, the expectation value of a random distribution of inclinations.
Note that the thin disk scale radius can be converted to a half-light radius
using the relation $r_{1/2} = 2.44\,r_{s}$.  The median of the microlensing
thin-disk size distribution is $\log (r_{s}/\textrm{cm}) = 16.2 \pm 0.5$, where the
error bar represents the bounds of the 68\% confidence interval.
When converted to a half-light radius, which
is comparable between different source models \citep{mortonson05},
the source size we obtain from our analysis [$\log (r_{1/2}/\textrm{cm}) = 16.5 \pm 0.5$] appears
to be marginally consistent with the $R$-band half-light radii (converted from $\sigma$
of a Gaussian disk profile) obtained by \citet[][$10^{15.5}\,\textrm{cm}$]{pelt98} and
\citet[][$< 10^{16.1}\,\textrm{cm}$]{refsdal00} from the data set
covering the 1982--1986 microlensing event and the following 8 years of stable flux ratio.
We note that our source size result is substantially larger than
those used by \citet{schmidt98} and \citet{wambsganss00} to constrain the microlens
mass distribution [$\log (r_{s}/\textrm{cm}) < 14.2$]; yet our mean microlens mass
is still consistent with their constraints.
We also experimented with the use of a uniform prior on the microlens mass of
$0.1 < \langle M/\msun \rangle < 1.0$.
Although the median of the resulting $r_{s}$ distribution is essentially the same
as that found without imposing the mass prior, within the uncertainties, the distribution yielded by the
prior contains a small but significant probability in solutions with small source sizes which
are not physically reasonable.  We suspect that the occurrence of this ``shelf"
at the low end of the size distribution is an artifact of a combination of the paucity of microlensing
variability in the observations and the mass prior itself.  By limiting $\langle M \rangle$
to a range about which its distribution (derived from the $v_{e}$ prior) is not
symmetric, we give more statistical weight to solutions from one side of the
distribution.  Since the Einstein velocity ($\hat{v}_{e}$) scale implied by a given microlens mass
scales as $\hat{v}_{e} = v_{e} \langle M/\msun \rangle^{-1/2}$, the solutions
with very low Einstein velocities are given more weight relative to the highest
velocity solutions when the prior is applied.  In the case of Q0957, the
solutions with low Einstein velocity correspond preferentially to small source sizes;
however, at very small sizes all solutions become equally likely since none
of them pass near a caustic.
  
\begin{figure}
\begin{center}
\plotone{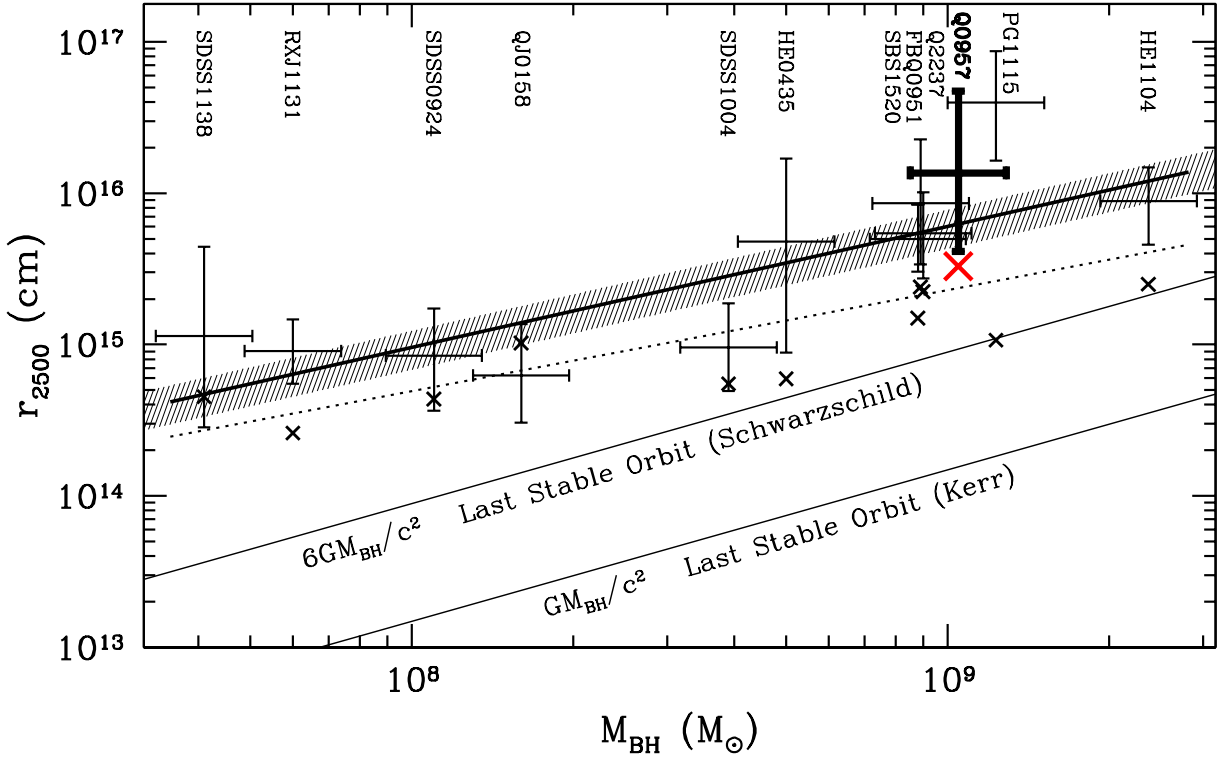}
\caption{Accretion disk size versus supermassive black hole mass relation 
(thick solid line) and data
from \citet{morgan10} including the new measurement of $r_{s}$ for Q0957, scaled to 
2500\,\AA\ and corrected to $60\degr$ inclination.  Q0957 is consistent with
the mean trend.  The dotted
line shows the scale radius as a function of central black hole mass predicted
by theoretical thin disk models (for $L/L_{E} = 1/3$ and $\eta = 0.1$), while the 
diagonal crosses indicate the thin disk size predicted by the magnification-corrected 
luminosity of the different quasars.  We find that Q0957 is
consistent with the findings of \citet{morgan10} and
\citet{blackburne11} that the optical continuum sizes of quasars measured
through microlensing analyses are larger than the sizes 
predicted by thin disk theory for a given black hole mass.  The microlensing
source size for Q0957 is also larger than the luminosity-constrained thin disk size, 
similar to the findings of \citet{pooley07}, \citet{morgan10} and 
\citet{mediavilla11}.}\label{fig:mbh_vs_r2500}
\end{center}
\end{figure}

\citet{morgan10} used the same microlensing analysis technique as we use here
on light curves of a sample of 11 different gravitationally lensed quasars to show that the
quasar accretion disk size at 2500\,\AA\ inferred from microlensing ($r_{2500}$) is 
correlated with central black hole mass
as $r_{2500} \propto M_{\textrm{BH}}^{0.80 \pm 0.17}$.  Although the scaling of the observed
correlation is consistent within the uncertainties with thin disk theory 
($R \propto M_{\textrm{BH}}^{2/3}$), the correlation
implies a quasar radiative efficiency $\eta$ that is approximately an order
of magnitude lower than is expected based on the local supermassive black hole mass and
quasar luminosity functions.  We can use this $r_{2500}$--$M_{\textrm{BH}}$ correlation
as an independent check on our assertion that the uncorrelated variability we have
observed in Q0957 is a result of microlensing.
To scale the observed-frame $r$-band source size we find for Q0957 to a rest-frame wavelength of 
2500\,\AA, we assume the $R\propto \lambda^{4/3}$ scaling of thin disk theory which
is supported by several observational studies of quasar microlensing 
\citep[e.g.,][]{anguita08,poindexter08,floyd09,mosquera11}. 
In doing so, we note that the effect of the wavelength scaling is minimal for Q0957 since 
the effective wavelength of the observed-frame $r$ filter corresponds to 2593\,\AA\ at $z=1.41$.
For the black hole mass we use the C\,\textsc{iv} emission-line estimate of \citet{assef11}, which has
an associated systematic uncertainty of 0.33 dex.   In Figure~\ref{fig:mbh_vs_r2500},
we place Q0957 on the best-fit $r_{2500}$--$M_{\textrm{BH}}$ relation found by \citeauthor{morgan10},
along with the quasar data points used to derive the correlation.  Q0957
is consistent with the relation; the agreement supports our assertion that  
microlensing is the source of the quasar's long-timescale uncorrelated variability.

Similarly to \citet{morgan10}, we also show for comparison in Figure~\ref{fig:mbh_vs_r2500}
quasar source sizes calculated by two additional methods.  First, we calculate the
theoretical scale radius of a thin accretion disk at a rest wavelength of 2500\,\AA\ 
from its central black hole mass (``theory size") using Eq.\ 2 of \citeauthor{morgan10},
and assuming accretion efficiency $\eta = 0.1$ and Eddington ratio 
$L/L_{E} = 1/3$ \citep{kollmeier06}.  Second, we calculate the disk size
under the thin disk model (a blackbody with a $T\propto R^{-3/4}$ temperature profile)  
constrained by the observed magnification-corrected optical flux/luminosity from \emph{HST} 
F814W measurements (``flux size", see Eq.\ 3 of \citeauthor{morgan10}).  
When we compare the results of the different methods of size calculation for Q0957, we observe
that the microlensing size is larger than the theory size,
also noted for other lensed quasars by \citet{morgan10} and \citet{blackburne11}.
\citet{pooley07} arrive at a similar conclusion, although they use bolometric
luminosity-based black hole masses to calculate their theory sizes
which are thus more similar to our flux sizes.  
Furthermore, the microlensing size for Q0957 is larger than its flux size as well, which was also
found by \citeauthor{morgan10}\ and \citet{mediavilla11} for their respective quasar samples.
However, the significance of the discrepancy we find for Q0957 is 
not as large as that typically found by \citet{blackburne11} for their sample: 
while they rule out the black hole mass-based theoretical prediction by at least $3\sigma$ in 
nearly every case, the flux size for Q0957 falls
within the 68\% confidence interval (essentially $1\sigma$) of the microlensing size,
and the theory size within the 80\% confidence interval ($\sim 1.3\sigma$).  
While this may simply reflect the relatively large
uncertainty in the microlensing size estimate for Q0957 compared to other 
microlensed quasars, it is also possible that the difference in significance is related to 
our different treatment of systematic errors from the analysis of
Blackburne et al.  The origin of the discrepancies between the 
microlensing size, theory size, and flux size remain unclear.  
As \citeauthor{morgan10}\ and \citeauthor{blackburne11}\ point out, adjusting the 
mass accretion rate ($L/\eta L_{E}$) of Q0957 by lowering $\eta$ or using
a higher-than-typical Eddington ratio may resolve the discrepancy of the
theory size with the microlensing size, but such adjustments will not address 
the fact that the flux size is smaller than the microlensing size. 

One possible explanation for the size discrepancy
is that the observed $r$-band flux may be contaminated by 
(a) UV/optical light from the continuum source scattered by the broad-line region (BLR);
or (b) higher energy continuum emission reprocessed by the BLR and re-emitted as
UV/optical emission lines.  Contamination from line emission
is an especially strong concern for our observations of Q0957 because the quasar's 
redshifted Mg\,\textsc{ii} line falls within the $r$ bandpass and 
could cause us to overestimate the microlensing size.  
Thus, it is important that we consider contamination scenarios in our analysis.  So,  
we have repeated our Monte Carlo light curve simulations assuming that different
percentages of the observed $r$-band flux can be attributed to 
unmicrolensed contamination from the BLR.
When we do so, we find that if we assume 10\% contamination 
we still obtain a median physical accretion disk size in $r$-band of 
$\log (r_{s, 10}/\textrm{cm}) = 16.1 \pm 0.5$.  
Even when we assume a 30\% contribution to the observed
flux from contamination, we obtain $\log (r_{s, 30}/\textrm{cm}) = 16.0^{+0.5}_{-0.4}$,
insignificantly different from our original microlensing size.  
Thus, neither scattering nor line contamination appear to produce
a significant effect on the microlensing size result, relative to our other
sources of uncertainty.  Moreover, even if line contamination produced a significant decrease in
the microlensing size, the flux size would be reduced 
along with the microlensing size, and the two would still not be reconciled.

Although we are not able to resolve the discrepancy between the different
accretion disk size estimates and our microlensing size measurement, we are nonetheless
very encouraged to observe a long-term microlensing event in a
quasar at the high end of the black hole mass function.
Moreover, Q0957 is relatively easy to monitor with ground-based observatories, since the quasar
images and lens galaxy are relatively widely separated and the time delay is 
well-determined.  Future multi-wavelength optical monitoring as well as X-ray 
monitoring will be most informative to examine the temperature profile of
the accretion disk and compare the X-ray and optical sizes.  

\acknowledgements

The authors wish to thank L.\ Goicoechea, E.\ Turner, \& R.\ McMillan for their assistance 
in obtaining historical photometric data for Q0957.  We also thank the anonymous referee
for suggestions which improved the data presented in this paper.
This material is based upon work supported by the National Science Foundation under 
grant No.\ AST-0907848 (to C.W.M.), AST-0708082, and AST-1009756 (to C.S.K.).
C.W.M. also gratefully acknowledges support from the Research 
Corporation for Science Advancement and Chandrasekhar X-Ray Center award 11700501.

\clearpage

\input{tab1}

\clearpage

\input{tab2}

\end{document}

%% file: tab1.tex
\begin{deluxetable}{lccccc}
\centering
\tablewidth{0pt}
\tablecolumns{6}
\tablecaption{\emph{HST} Astrometry and Photometry of Q\,0957+561\label{tab:q0957_astrom}}
\tablehead{
\colhead{} & \multicolumn{2}{c}{Astrometry} & \multicolumn{3}{c}{Photometry} \\ 
\colhead{Component} & \colhead{$\Delta\textrm{R.A.}$} & \colhead{$\Delta\textrm{Dec.}$} & 
    \colhead{F555W} & \colhead{F814W} & \colhead{F160W} \\
\colhead{} & \colhead{(arcsec)} & \colhead{(arcsec)} & \colhead{(mag)} & \colhead{(mag)} & \colhead{(mag)} 
}
\startdata
A & $\equiv 0$ & $\equiv 0$ & $17.09 \pm 0.07$ & $16.71 \pm 0.05$ & $15.60 \pm 0.03$ \\
B & $+1.229 \pm 0.005$ & $-6.048 \pm 0.004$ & $17.11 \pm 0.05$ & $16.78 \pm 0.04$ & $15.68 \pm 0.03$ \\
Lens Galaxy & $+1.406 \pm 0.006$ & $-5.027 \pm 0.005$ & $19.05 \pm 0.04$ & $17.12 \pm 0.02$ & $15.14 \pm 0.09$ \\
\enddata
\tablecomments{\emph{HST} astrometry and photometry for Q0957 are taken from \citet{keeton00}. }

\end{deluxetable}

%% file: tab2.tex
\begin{deluxetable}{lcccc}
\centering
\tabletypesize{\small}
\tablewidth{0pt}
\tablecolumns{5}
\tablecaption{Q\,0957+561 Light Curves \label{tab:lightcurve}}
\tablehead{
\colhead{HJD - 2450000} & \colhead{Seeing} & \colhead{QSO A} & \colhead{QSO B} & \colhead{$\langle \textrm{Stars} \rangle$} \\
\colhead{(days)}         & \colhead{(arcsec)} & \colhead{(mag)} & \colhead{(mag)} & \colhead{(mag)}
}
\startdata
$4554.712$ & $1.3$ & $ 2.671\pm 0.004$ & $ 2.419\pm 0.004$ & $ 0.013\pm 0.002$ \\ 
$4555.767$ & $1.4$ & $ 2.659\pm 0.004$ & $ 2.412\pm 0.004$ & $ 0.010\pm 0.002$ \\ 
$4561.722$ & $1.3$ & $ 2.665\pm 0.004$ & $ 2.417\pm 0.004$ & $ 0.013\pm 0.002$ \\ 
$4570.669$ & $1.5$ & $ 2.669\pm 0.004$ & $ 2.413\pm 0.004$ & $-0.004\pm 0.002$ \\ 
$4584.677$ & $1.4$ & $ 2.664\pm 0.004$ & $ 2.378\pm 0.004$ & $ 0.008\pm 0.002$ \\ 
$4596.704$ & $1.4$ & $ 2.682\pm 0.004$ & $ 2.377\pm 0.005$ & $-0.002\pm 0.002$ \\ 
$4613.691$ & $1.4$ & $ 2.689\pm 0.004$ & $ 2.392\pm 0.004$ & $ 0.008\pm 0.002$ \\ 
$4617.702$ & $1.3$ & $ 2.691\pm 0.004$ & $ 2.412\pm 0.004$ & $ 0.010\pm 0.002$ \\ 
$4793.005$ & $0.9$ & $ 2.733\pm 0.004$ & $ 2.541\pm 0.005$ & $ 0.007\pm 0.002$ \\ 
$4802.980$ & $1.4$ & $ 2.711\pm 0.005$ & $ 2.544\pm 0.005$ & $-0.008\pm 0.002$ \\ 
$4807.929$ & $1.3$ & $ 2.728\pm 0.005$ & $ 2.563\pm 0.005$ & $-0.002\pm 0.002$ \\ 
$4829.857$ & $1.3$ & $ 2.747\pm 0.004$ & $ 2.604\pm 0.004$ & $ 0.007\pm 0.002$ \\ 
$4833.894$ & $1.4$ & $ 2.726\pm 0.004$ & $ 2.603\pm 0.004$ & $ 0.006\pm 0.002$ \\ 
$4839.966$ & $1.6$ & $ 2.722\pm 0.005$ & $ 2.608\pm 0.005$ & $-0.017\pm 0.002$ \\ 
$4846.758$ & $1.8$ & $ 2.730\pm 0.004$ & $ 2.621\pm 0.004$ & $-0.011\pm 0.002$ \\ 
$4862.882$ & $1.2$ & $ 2.706\pm 0.004$ & $ 2.631\pm 0.004$ & $ 0.013\pm 0.002$ \\ 
$4877.716$ & $2.0$ & $ 2.684\pm 0.004$ & $ 2.618\pm 0.004$ & $-0.009\pm 0.002$ \\ 
$4883.932$ & $1.1$ & $ 2.673\pm 0.002$ & $ 2.624\pm 0.002$ & $-0.078\pm 0.002$ \\ 
$4891.824$ & $1.3$ & $ 2.635\pm 0.004$ & $ 2.625\pm 0.004$ & $ 0.054\pm 0.001$ \\ 
$4911.800$ & $1.1$ & $ 2.618\pm 0.004$ & $ 2.665\pm 0.004$ & $ 0.013\pm 0.002$ \\ 
$4942.797$ & $1.7$ & $ 2.573\pm 0.003$ & $ 2.655\pm 0.004$ & $ 0.011\pm 0.001$ \\ 
$4964.715$ & $1.2$ & $ 2.587\pm 0.004$ & $ 2.630\pm 0.004$ & $ 0.013\pm 0.001$ \\ 
$4981.695$ & $2.2$ & $ 2.580\pm 0.005$ & $ 2.608\pm 0.005$ & $-0.028\pm 0.002$ \\ 
$4997.677$ & $1.2$ & $ 2.608\pm 0.004$ & $ 2.626\pm 0.004$ & $ 0.013\pm 0.002$ \\ 
$5144.981$ & $1.3$ & $ 2.573\pm 0.004$ & $ 2.656\pm 0.004$ & $ 0.004\pm 0.002$ \\ 
$5157.964$ & $1.1$ & $ 2.578\pm 0.004$ & $ 2.642\pm 0.004$ & $ 0.001\pm 0.002$ \\ 
$5186.024$ & $1.2$ & $ 2.558\pm 0.004$ & $ 2.603\pm 0.004$ & $ 0.005\pm 0.002$ \\ 
$5201.946$ & $1.2$ & $ 2.579\pm 0.004$ & $ 2.639\pm 0.004$ & $ 0.003\pm 0.002$ \\ 
$5208.993$ & $1.2$ & $ 2.585\pm 0.004$ & $ 2.652\pm 0.004$ & $ 0.009\pm 0.002$ \\ 
$5241.874$ & $2.1$ & $ 2.627\pm 0.004$ & $ 2.668\pm 0.004$ & $-0.010\pm 0.002$ \\ 
$5261.687$ & $1.4$ & $ 2.631\pm 0.004$ & $ 2.667\pm 0.005$ & $-0.016\pm 0.002$ \\ 
$5296.691$ & $0.8$ & $ 2.641\pm 0.004$ & $ 2.617\pm 0.004$ & $ 0.015\pm 0.002$ \\ 
$5311.690$ & $1.3$ & $ 2.634\pm 0.004$ & $ 2.575\pm 0.005$ & $-0.011\pm 0.002$ \\ 
$5320.745$ & $0.9$ & $ 2.639\pm 0.004$ & $ 2.546\pm 0.004$ & $ 0.017\pm 0.001$ \\ 
$5324.729$ & $0.7$ & $ 2.636\pm 0.004$ & $ 2.536\pm 0.004$ & $ 0.018\pm 0.001$ \\ 
$5332.717$ & $1.1$ & $ 2.630\pm 0.004$ & $ 2.528\pm 0.004$ & $ 0.065\pm 0.002$ \\ 
$5348.695$ & $1.1$ & $ 2.605\pm 0.004$ & $ 2.480\pm 0.004$ & $ 0.013\pm 0.001$ \\ 
$5352.685$ & $0.9$ & $ 2.596\pm 0.004$ & $ 2.483\pm 0.004$ & $ 0.014\pm 0.002$ \\ 
$5358.688$ & $1.8$ & $ 2.560\pm 0.004$ & $ 2.463\pm 0.005$ & $-0.003\pm 0.002$ \\ 
$5506.991$ & $1.4$ & $ 2.650\pm 0.004$ & $ 2.474\pm 0.004$ & $ 0.007\pm 0.002$ \\ 
$5519.030$ & $0.9$ & $ 2.712\pm 0.003$ & $ 2.490\pm 0.003$ & $-0.008\pm 0.001$ \\ 
$5527.937$ & $1.7$ & $ 2.692\pm 0.004$ & $ 2.483\pm 0.004$ & $-0.008\pm 0.002$ \\ 
$5535.032$ & $1.1$ & $ 2.704\pm 0.006$ & $ 2.489\pm 0.006$ & $-0.016\pm 0.002$ \\ 
$5543.989$ & $1.0$ & $ 2.713\pm 0.004$ & $ 2.497\pm 0.004$ & $ 0.014\pm 0.002$ \\ 
$5563.958$ & $1.3$ & $ 2.705\pm 0.004$ & $ 2.477\pm 0.004$ & $ 0.009\pm 0.001$ \\ 
$5590.779$ & $1.0$ & $ 2.729\pm 0.004$ & $ 2.496\pm 0.004$ & $ 0.017\pm 0.001$ \\ 
$5604.847$ & $1.5$ & $ 2.711\pm 0.004$ & $ 2.465\pm 0.004$ & $ 0.004\pm 0.002$ \\ 
$5621.687$ & $1.0$ & $ 2.705\pm 0.004$ & $ 2.482\pm 0.004$ & $ 0.012\pm 0.002$ \\ 
$5626.779$ & $1.2$ & $ 2.698\pm 0.005$ & $ 2.487\pm 0.005$ & $ 0.003\pm 0.002$ \\ 
$5649.715$ & $1.1$ & $ 2.661\pm 0.004$ & $ 2.538\pm 0.004$ & $ 0.015\pm 0.001$ \\ 
$5653.710$ & $1.1$ & $ 2.659\pm 0.004$ & $ 2.545\pm 0.004$ & $ 0.014\pm 0.001$ \\ 
$5664.750$ & $1.2$ & $ 2.658\pm 0.004$ & $ 2.559\pm 0.004$ & $ 0.003\pm 0.002$ \\ 
$5674.748$ & $1.6$ & $ 2.648\pm 0.008$ & $ 2.566\pm 0.008$ & $-0.091\pm 0.003$ \\ 
$5684.672$ & $1.3$ & $ 2.653\pm 0.004$ & $ 2.564\pm 0.004$ & $-0.007\pm 0.001$ \\ 
$5702.677$ & $1.7$ & $ 2.663\pm 0.004$ & $ 2.572\pm 0.004$ & $ 0.005\pm 0.002$ \\ 
$5712.673$ & $1.0$ & $ 2.671\pm 0.004$ & $ 2.582\pm 0.004$ & $ 0.012\pm 0.002$ \\ 
$5733.651$ & $1.2$ & $ 2.673\pm 0.005$ & $ 2.566\pm 0.005$ & $-0.065\pm 0.002$ \\ 
\enddata
\tablecomments{HJD is the Heliocentric Julian Day.  The magnitudes listed in the QSO 
A and B columns are measured relative to the comparison stars.  
The magnitudes in the $\langle \textrm{Stars} \rangle$
column are the mean magnitudes of the comparison stars for that epoch relative to
their mean over all epochs.}

\end{deluxetable}